\documentclass{article}
\usepackage{graphics,amsmath,mathrsfs,amsfonts,amsthm}
\usepackage{picins}
\usepackage{graphicx}
\usepackage{subfigure}
\begin{document}

\title{\centering \textbf{ A novel efficient numerical solution of Poisson equation for arbitrary shapes in two dimensions}}

\author{Zu-Hui Ma\footnote{Department of Electric and Electronic Engineering, The University of Hong Kong, Hong Kong (mazuhui@hku.hk)} , Weng Cho Chew\footnote{Corresponding author. Department of Electrical and Computer Engineering, University of Illinois at Urbana-Champaign, Urbana, IL 61801 USA (w-chew@uiuc.edu)} and Li Jun Jiang\footnote{Department of Electric and Electronic Engineering, The University of Hong Kong, Hong Kong (ljiang@eee.hku.hk)}}

\date{}

\bigskip
\medskip
\maketitle
\begin{abstract}
We propose a novel efficient algorithm to solve Poisson equation in
irregular two dimensional domains for electrostatics. It can handle
Dirichlet, Neumann or mixed boundary problems in which the filling
media can be homogeneous or inhomogeneous.

The basic idea of the new method is solve the problem in three
steps: (i) First solve the equation $\nabla\cdot\mathbf D=\rho$. The
inverse of the divergence operator in a restricted subspace is found
to yield the electric flux density $\mathbf D$ by a fast direct
solver in $O(N)$ operations. The $\mathbf D$ so obtained is
nonunique with indeterminate divergence-free component. Then the
electric field is found by $\mathbf E=\mathbf D/\epsilon$. But
$\nabla\times\mathbf E=0$ for electrostatic field; hence, $\mathbf
E$ is curl free and orthogonal to the divergence free space. (ii) An
orthogonalization process is used to purify the electric field
making it curl free and unique. (iii) Then the potential $\phi$ is
obtained by solving $\nabla \phi=-\mathbf E$ or finding the inverse
of the gradient operator in a restricted subspace by a similar fast
direct solver in $O(N)$ operations.

Treatments for both Dirichlet and Neumann boundary conditions are
addressed. Finally, the validation and efficiency are illustrated by
several numerical examples. Through these simulations, it is
observed that the computational complexity of our proposed method
almost scales as $\textit{O}(N)$, where $N$ is the triangle patch
number of meshes. Consequently, this new algorithm is a feasible
fast Poisson solver.

\end{abstract}


\bigskip
\bigskip
\bigskip
\bigskip

\section{Introduction}
\label{}

The Poisson equation occurs in the analysis and modeling of many
scientific and engineering problems. In electrostatics, Poisson
equation arises when finding the electrostatic potential of an
electric field in a region with continuously distributed charges
\cite{Jackson_EM_book}. It is often solved in micro- and
nano-electronic device physics \cite{Datta_book}, as well as in
electronic transport and electrochemistry problems in terms of the
Poisson-Boltzmann equation \cite{Pois_BoltzEqu}. In fluid dynamics,
Poisson equations are solved to find the velocity potential in a
steady-state potential flow of an incompressible fluid  with
internal sources or sinks \cite{ElmanFemBook}. Moreover, Poisson
equation is also encountered in finding the steady-state temperature
in an isotropic body with internal sources \cite{FeynmanLecture2}.

An accurate and efficient solution of Poisson equation is critical
in various areas. For example, in design optimization of nanodevices
where quantum effects are significant, a widely used scheme is to
solve the coupled Schr\"{o}dinger-Poisson system self-consistently
\cite{ChengLiu2007,HuangChew2012}, in which Poisson equation is
solved repeatedly, and concomitantly with the Schr\"{o}dinger's
equation. As such, the computational load for solving Poisson
equation is always of concern.

There are two main classes of solvers for linear systems from
Poisson equation: direct and iterative. One of the direct solvers is
the fast Poisson solver based on fast Fourier transform (FFT)
\cite{PSfft}. Indeed, this method is extremely efficient when the
solution regions are simple and regular geometries with regular
grids, such as rectangular regions, 2-D polar and spherical
geometries \cite{LaiPS2002}, and spherical shells
\cite{HuangFFT2011}. Since practical problems usually involve
complex geometries, there have been many research works on seeking
alternative methods. The multifrontal method with nested dissection
ordering \cite{Davis_MF1997} is the most efficient direct method
that can deal with complex geometries. Its key idea relies on
partitioning the domain using a nested hierarchical structure and
generating the $LU$ decomposition from bottom up to minimize
fill-ins. Typically, the computational complexity of the
multifrontal method is of $O(N^{1.5})$ in two dimensions where $N$
is the dimension of the matrix.

The other class of solvers, the iterative ones, are more favorable
when large systems are solved. A popular one is the approach based
on integral equation techniques and accelerated by fast multipole
method (FMM)
\cite{McKenney1995,HuangGreengard2000,Ethridge2001,Langston2011}. It
can achieve $\textit{O}(N)$ complexity when the underlying Green's
function is available and amenable to factorization. Nevertheless,
the underlying Green's function is difficult to obtain unless
involved problems have constant coefficient in free-space or
separate geometries with simple boundary conditions.

Multigrid method is one of most effective preconditioning strategies
for iterative Poisson solvers \cite{MG_book,MG_rev,MGtutorial2000}.
Although this method takes advantage of fine meshes and coarse
meshes and can achieve nearly optimal efficiency in theory, it is
difficult to implement in a robust fashion because it demands a set
of hierarchical grids of different density, which is not convenient
in many real world problems.

In this paper, we present a new efficient numerical solution of
Poisson equation for arbitrary two-dimensional domain with
homogeneous or inhomogeneous media. Unlike traditional Poisson
solvers, where the potential is found directly, this new method
solves for electric flux density $\mathbf D$ first. The electric
flux\footnote{The ``electric flux density" is also called the
electric displacement field.  We will call it ``electric flux" for
short.} is first approximated by the RWG basis. It uses the
loop-tree decomposition of the RWG approximated vectorial electric
flux which is a quasi-Helmholtz decomposition method developed in
computational electromagnetics (CEM). With this technique, the
electric flux is expressed as the combination of the loop space
(subspace) (solenoidal or divergence free) part and the tree space
(subspace) (quasi-irrotational) part.  These two spaces, however,
are non-orthogonal to each other, unlike the case of a pure
Helmholtz decomposition.

Expanding the electric flux with two sets of basis functions: loop
and tree basis functions, we can solve for the electric flux by a
two-stage process: First, to find the tree-space part, a matrix
system is derived from the differential equation,
$\nabla\cdot\mathbf{D}=\rho$, and then solved by a fast tree direct
solver with $\textit{O}(N_t)$ complexity, where $N_t$ is the total
number of tree basis functions.  The obtained electric flux is
nonunique as a divergence free component can always be added to it.

Second, the electric field $\mathbf E=\mathbf D/\epsilon$ is
obtained but it is incorrect because $\mathbf E$ should be curl-free
but the so-obtained electric field is not curl-free.  It is polluted
by the loop-space components.
The loop-space part of $\mathbf E$ is purged by a projection
procedure, which is iterative.
Having acquired the curl-free
electric field distribution, which is unique, we can readily get the
potential distribution by solving $\mathbf E=-\nabla \phi$ by the
fast tree solver as well.

In addition, we address the special treatments for Dirichlet and
Neumann boundary conditions, respectively, which guarantee that the
obtained solution is unique. Through numerical examples, we validate
the feasibility and efficiency of the new method whose computational
complexity almost scales as $\textit{O}(N)$, where $N$ is the
triangle patch number of meshes.

To the best of our knowledge, this is the first time that
quasi-Helmholtz decomposition techniques, e.g., loop-tree
decomposition, are introduced to solve Poisson equation. This new
method provides a feasible alternative for existing fast Poisson
solvers. Moreover, the work about Poisson equation with Neumann
boundary condition in homogeneous media has been reported
in~\cite{mySISC}. Here, not only does this paper extend the method
to inhomogeneous medium cases, but it also addresses the special
treatment for the Dirichlet boundary condition.

The organization of this paper is as follows. In
Section~\ref{Sec:S2}, we introduce the basic problem definition and
several relevant preliminaries. In Section~\ref{Sec:S3}, we present
the algorithm for solving Poisson problem with Neumann boundary
condition. Next, we offer a special treatment for Dirichlet boundary
condition in Section~\ref{Sec:S4}. Finally, in Section~\ref{Sec:S5},
we will verify the validation and illustrate the efficiency of the
new method by several numerical methods. Conclusions will be drawn
in Section~\ref{Sec:S6}.

\section{Preliminaries}\label{Sec:S2}

\subsection{Two dimensional Poisson problems}
\label{}
\begin{figure}[h]
\centerline{\includegraphics[width=0.4\columnwidth,draft=false]{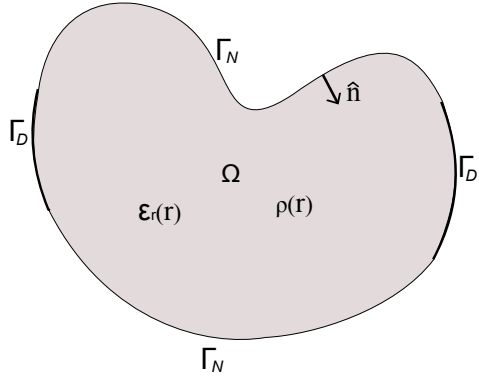}}
\caption{Schema for a typical Poisson problem.} \label{fig:PE}
\end{figure}

Assume inhomogeneous dielectric materials occupying a two
dimensional bounded and simply connected region, $\Omega$, with
boundary $\Gamma$ and normal $\hat{\mathbf{n}}$ that points to the
solution region as shown in Fig.~\ref{fig:PE}. The boundary $\Gamma$
is composed of two parts: the first one, denoted by $\Gamma_D$, is
imposed by Dirichlet boundary condition and the other part,
$\Gamma_N$, is imposed by Neumann boundary condition.

In this paper, we are interested in solving the 2-D Poisson equation arising from electrostatic problem:
\begin{equation}\label{s2eq1}
\renewcommand{\arraystretch}{1.5}
\begin{array}{ll}
\nabla\cdot\epsilon_r(\mathbf{r})\nabla\phi(\mathbf{r})=-\rho(\mathbf{r})/\epsilon_0 &\mbox{ for $\mathbf{r}\in\Omega$}   \\
\phi(\mathbf{r})=\phi_0(\mathbf{r}) &\mbox{ for $\mathbf{r}\in\Gamma_D$}  \\
\frac{\partial}{\partial n}\phi(\mathbf{r})=g(\mathbf{r}) &\mbox{ for $\mathbf{r}\in\Gamma_N$}
\end{array}
\end{equation}
where $\phi$ is the potential and $\rho$ describes the charge
density, $\epsilon_r$ and $\epsilon_0$ are the relative permittivity
and free space permittivity, respectively. Suppose that the
Dirichlet boundary consists of finite $M$ distinct boundaries,
$\Gamma_D= \bigcup_{i=1}^{M}\Gamma_D^{(i)}$, then a fixed potential
$\phi_0^{(i)}$ is prescribed on the boundary $\Gamma_D^{(i)}$ for
$i=1,2,...,M$. Moreover, to complete the description of a well-posed
problem, the Neumann boundary data $g(\mathbf{r})$ must be a square
integrable function over the corresponding boundary
\cite{Brown1994}.

Apparently, when $\Gamma_N=0$, the above equation shrinks to a
Dirichlet problem, which has unique solution. Similarly, it becomes
a Neumann problem that is uniquely solvable (up to a constant) when
$\Gamma_D=0$.

\subsection{RWG basis}
\begin{figure}[h]
\centerline{\includegraphics[width=0.4\columnwidth,draft=false]{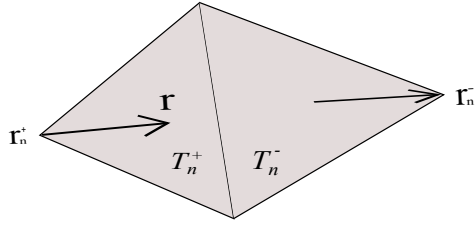}}
\caption{The geometry description of a Rao-Wilton-Glisson (RWG) function.}
\label{fig:rwg}
\end{figure}
The Rao-Wilton-Glisson (RWG) function \cite{RWG} is the most popular
choice of expansion function (also called basis function) in the
computational electromagnetics (CEM) community. In this paper, we
use a normalized version by removing the edge length from the
original definition. As shown in Fig.~\ref{fig:rwg}, the normalized
RWG function straddles two adjacent triangles, and hence, its
description requires two contiguous triangles. The expression for
the expansion function describing the field on the triangle is
\begin{equation}
\mathbf{\Lambda}_n(\mathbf{r})= \left\{
\renewcommand{\arraystretch}{1.5}
 \begin{array}{cl}
 \pm\frac{1}{2A_n^{\pm}}(\mathbf{r}-\mathbf{r}_n^{\pm}) &\mbox{ if $\mathbf{r}\in T_n^{\pm}$} \\
  0 &\mbox{ otherwise}
 \end{array} \right.
\end{equation}
where $\pm$ denote the respective triangles, $\mathbf{r}_n^{\pm}$
and $A_n^{\pm}$ are the vertex points and areas of the respective
triangles,  and $T_n^{\pm}$ are the supports of the respective
triangles.

Apparently, RWG function is the lowest order divergence conforming
function which is defined on a pair of triangles. Such basis
functions can be used to expand the electric flux
\begin{equation}
\mathbf{D}(\mathbf{r})=\sum_{i=1}^{N}\mathbf{\Lambda}_n(\mathbf{r})
\end{equation}
where $N$ is the total number of normalized RWG basis functions. Here, $\mathbf{D}$ is defined by
\begin{equation}
\mathbf{D}(\mathbf{r})=-\epsilon(\mathbf{r})\nabla\phi(\mathbf{r})
\end{equation}
where
\begin{equation}
\epsilon(\mathbf{r})=\epsilon_0\epsilon_r(\mathbf{r})
\end{equation}
is the electric permittivity.

The divergence of $\mathbf{\Lambda}_n$, which is proportional to the charge density associated the basis function, reads
\begin{equation}
\nabla\cdot\mathbf{\Lambda}_n(\mathbf{r})= \left\{
\renewcommand{\arraystretch}{1.5}
 \begin{array}{cl}
 \pm\frac{1}{A_n^{\pm}} &\mbox{ if $\mathbf{r}\in T_n^{\pm}$} \\
  0 &\mbox{ otherwise}.
       \end{array} \right.
\end{equation}
The charge density is thus constant in each triangle. The total
charge associated the triangle pair, $T_n^+$ and $T_n^-$, is zero.
This implies charge neutrality physically. Hence, the divergence
conforming property of RWG function makes its representation of the
field physical.

\subsection{Helmholtz theorem and loop-tree decomposition}
The well-known Helmholtz theorem \cite{BladelEMbook} states that a vector field can be split into the form
\begin{equation}
\mathbf{f}(\mathbf{r})=\nabla\varphi+\nabla\times\mathbf{v}.
\end{equation}
The first term $\nabla\varphi$ is the irrotational (curl-free) part,
and the second term $\nabla\times\mathbf{v}$ is the solenoidal
(divergence-free) part. Low frequency problem has been an intense
research area for the last decade in the CEM community. When the
frequency is low, the current naturally decomposes into a solenoidal
(divergence-free) part and an irrotational (curl-free) part. These
two parts are  imbalanced when the frequency becomes low. Hence,
there is a severe low frequency breakdwown numerical problem when
solving integral equations in which RWG function is normally used.
One remarkable remedy is the well known loop-tree decomposition
\cite{Wilton1981,Mautz1984,JSZhao2000,Wu1994,Burton1995,ChewBook2008},
in which RWG basis is decomposed into the loop basis functions that
have zero divergence, and the tree basis (or star basis) functions
that have nonzero divergence. This is a quasi-Helmholtz
decomposition because the tree expansion functions are not curl
free. For short, we call the subspace spanned by RWG basis functions
the RWG space, the subspace spanned by the tree basis functions the
tree space, and that spanned by loop basis functions the loop space.

Quite similar to the case of CEM, we can expand the electric flux
\begin{equation}\label{Eq:lpd}
\mathbf{D}(\mathbf{r})=\mathbf{D}_l(\mathbf{r})+\mathbf{D}_t(\mathbf{r})=\sum_{i=1}^{N_l}
l_i\mathbf{L}_i(\mathbf{r})+\sum_{i=1}^{N_t}
t_i\mathbf{T}_i(\mathbf{r})
\end{equation}
where $\mathbf{D}_l$ and $\mathbf{D}_t$ are the loop-space part and
the tree space part respectively, $\mathbf{L}_i(\mathbf{r})$ is a
loop expansion function such that
$\nabla\cdot\mathbf{L}_i(\mathbf{r})=0$, and
$\mathbf{T}_i(\mathbf{r})$ is a tree expansion function such that
$\nabla\cdot\mathbf{T}_i(\mathbf{r})\neq 0$, the numbers of the loop
basis functions and the tree basis functions are $N_l$ and $N_t$,
respectively.

\subsection{Loop basis and tree basis}

\begin{figure}[h]
  \centering
  \subfigure{
    \label{fig:loop:a} 
    \includegraphics[width=1.4in]{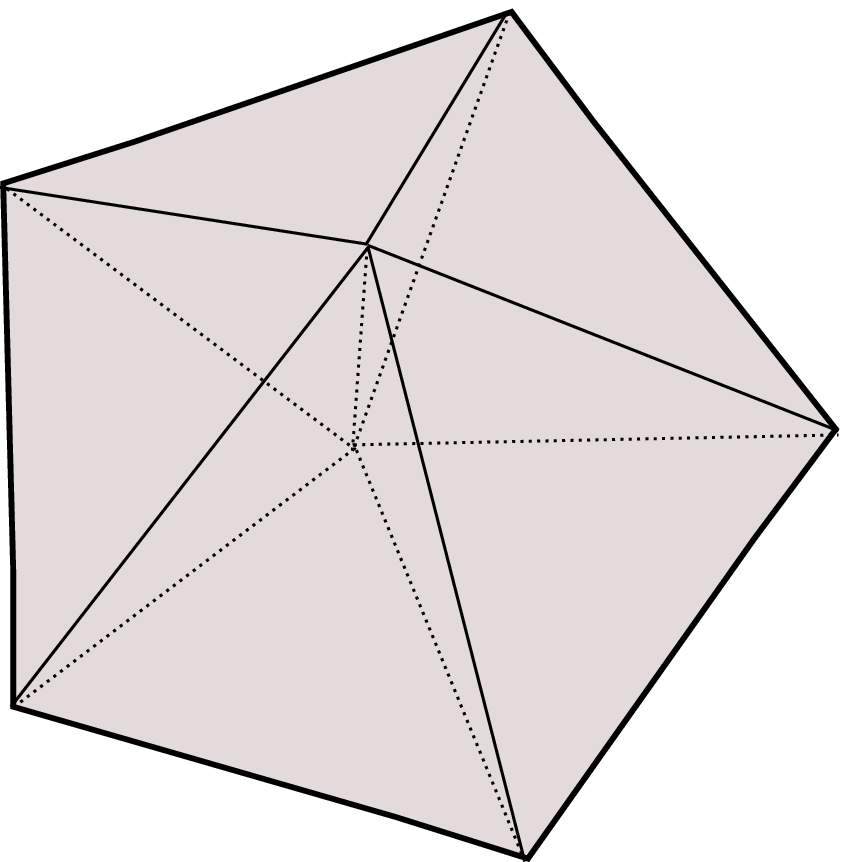}}
  \hspace{1in}
  \subfigure{
    \label{fig:loop:b} 
    \includegraphics[width=1.4in]{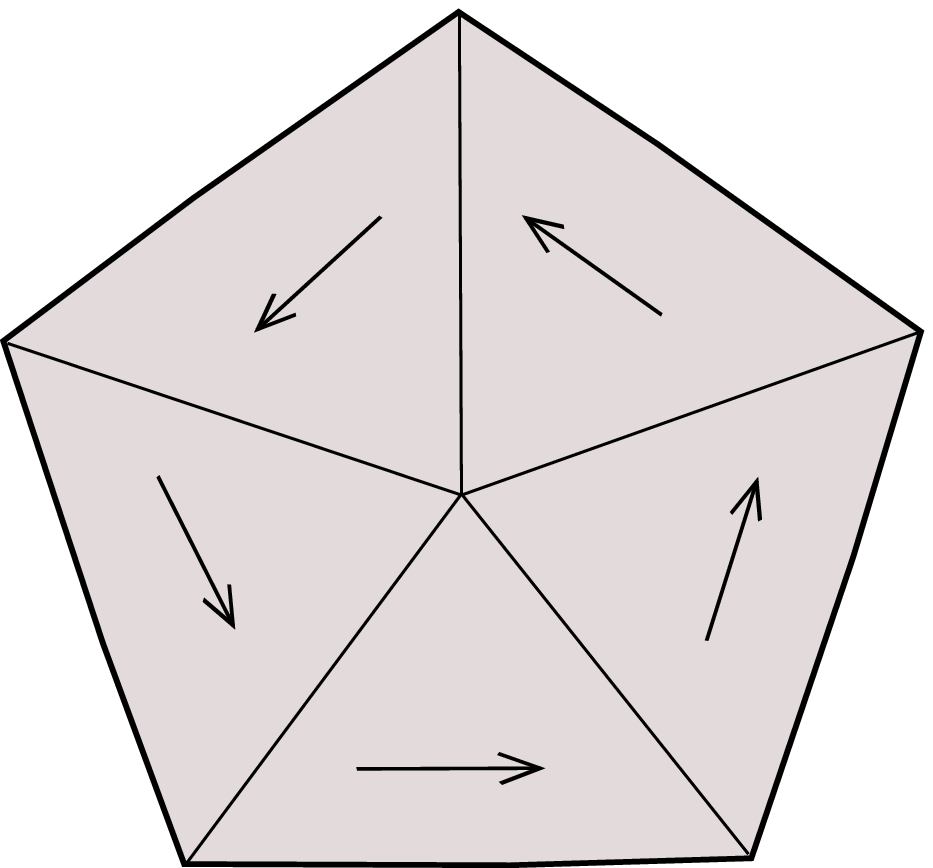}}
  \caption{Linear pyramid basis function and loop basis function. Left: pyramid function. Right: loop basis function.}
  \label{fig:loop} 
\end{figure}

A loop basis function is described by the surface curl of a vector function, namely,
\begin{equation}
\mathbf{L}(\mathbf{r})=\nabla_s\times\hat{z}\,\Delta(\mathbf{r})
\end{equation}
where the scalar function $\Delta(\mathbf{r})$, also referred as
``solenoidal potential" \cite{VecchiLS}, is the linear Lagrange or
nodal interpolating basis. As shown in the left part of
Fig.~\ref{fig:loop}, $\Delta_i(\mathbf{r})$ is a piecewise linear
function with support on the triangles that have a vertex at the
$i$th node of the mesh, attaining a unit value at node $i$, and
linearly approaching zero on all neighboring nodes. This
interpolating scalar functions is also intuitively called pyramid
basis functions. Moreover, the right one of Fig.~\ref{fig:loop}
illustrates the loop basis function $\mathbf{L}_i$ associated with
an interior node $i$. Within the triangles attached to node $i$,
$\mathbf{L}_i$ has a vector direction parallel to the edge opposite
to node $i$ and forms a loop around node $i$.

\begin{figure}[h]
\centerline{\includegraphics[width=0.3\columnwidth,draft=false]{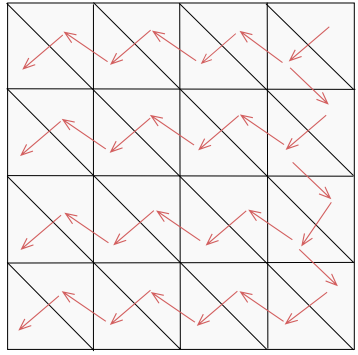}}
\caption{A possible tree basis illustration. Every arrow stands for
an RWG function that is the member of tree basis.} \label{fig:tree}
\end{figure}

One the other hand, the tree basis consists of RWG functions that
lie along a tree structure connecting the centroids of adjacent
triangular patches. Fig.~\ref{fig:tree} shows one possible choice of
the tree basis for this particular triangular mesh.

\section{Proposed method for Neumann problems}\label{Sec:S3}

For Neumann problems, the equation in question becomes
\begin{equation}
\renewcommand{\arraystretch}{1.5}
\begin{array}{ll}
\displaystyle \nabla\cdot\epsilon_r(\mathbf{r})\nabla\phi(\mathbf{r})=-\rho(\mathbf{r})/\epsilon_0 &\mbox{ for $\mathbf{r}\in\Omega$} \\
\frac{\partial}{\partial n}\phi(\mathbf{r})=g(\mathbf{r}) &\mbox{ for $\mathbf{r}\in\Gamma$}
\end{array}
\end{equation}
where $\Gamma$ includes all the boundaries of $\Omega$.

A necessary condition for the existence of a solution to this
problem is that the source and boundary data satisfy the
compatibility condition \cite{HansonBook}
\begin{equation}\label{s3eq2.1}
\int_{\Gamma} -\epsilon(\mathbf{r}) g(\mathbf{r}) d\textit{l}+ \int_{\Omega} \rho(\mathbf{r}) d\mathbf{r} =0.
\end{equation}

For Neumann problems in homogeneous media, the basic idea of
proposed method and the solution process have been reported
in~\cite{mySISC}. In this section, we extend this approach to handle
more complex Poisson problems that involve inhomogeneous media.

\subsection{Algorithm outline}\label{S3}
We outline the overall solution procedures as follows:

\textbf{1. Acquire the tree space part of electric flux density:} By
solving a matrix system resulting from $\nabla\cdot\mathbf{D}=\rho$,
we first obtain the tree space part of electric flux density,
$\mathbf{D}_{t}$, that corresponds to the second term on the right
side of Eq.~(\ref{Eq:lpd}). Taking advantage of the fast tree
solver, we can obtain $\mathbf{D}_{t}$ in $\textit{O}(N_t)$
operations.

\textbf{2. Field projection onto loop space:} In order to get the
first term on the right side of Eq.~(\ref{Eq:lpd}), the loop-space
part $\mathbf{D}_{l}$, we need to follow up with a projection
procedure. Since the electric field $\mathbf E$ is curl free and
orthogonal to the divergence-free space, this can be achieved by
projecting the electric field $\mathbf{E}$ onto the loop space and
orthogonalizing it with respect to the loop space, as presented in
Section~\ref{S3s3}. By this technique, the desired electric field
will be retrieved.

\textbf{3. Solution of potential:} The potential distribution can be
found by solving another matrix system accelerated by the fast tree
solver that is of $\textit{O}(N_t)$, in a manner similar to that of
the first procedure.

\subsection{Solution of $\mathbf{D}_t$}\label{S3s2}

We start with the loop-tree decomposition of electric flux
\begin{equation}\label{s3eq3}
\mathbf{D}(\mathbf{r})=\sum_{i=1}^{N_l} l_i\mathbf{L}_i(\mathbf{r})+\sum_{i=1}^{N_t} t_i\mathbf{T}_i(\mathbf{r})
\end{equation}
where the first term of right hand side is the loop-space part, and
the second one is the tree space part.

Meanwhile, from electrostatic theory, we have
\begin{equation}\label{s3eq5}
\nabla\cdot\mathbf{D}(\mathbf{r})=\rho(\mathbf{r}).
\end{equation}
The charge density $\rho(\mathbf{r})$ can be expanded in terms of pulse expansion functions \cite{ChewBook2008}, namely,
\begin{equation}\label{s3eq6}
\rho(\mathbf{r})=\sum_{i=1}^{N_p} q_n P_n(\mathbf{r})
\end{equation}
where $N_p$ is the number of triangle patches.

When Eqs.~(\ref{s3eq3}) and (\ref{s3eq6}) are substituted into
Eq.~(\ref{s3eq5}) and testing Eq.~(\ref{s3eq5}) with a set of pulse
functions, we have a matrix system
\begin{equation}\label{s3eq9}
\overline{\mathbf{K}}\cdot\textbf{I}_t= \textbf{V}_{\rho}
\end{equation}
where
\begin{equation}\label{s3eq10}
\left[\overline{\mathbf{K}}\right]_{ij}= \left\langle P_i(\mathbf{r}),\nabla \cdot \textbf{T}_j(\mathbf{r}) \right\rangle
\end{equation}
\begin{equation}\label{s3eq11}
\left[\textbf{V}_{\rho}\right]_i= \left\langle P_i(\mathbf{r}),\rho(\mathbf{r})\right\rangle = q_i
\end{equation}
and $\textbf{I}_t=[t_1\;t_2\;t_3\;\cdots\; t_{N_t}]^T$. The number
of tree basis functions, $N_t$, is equal to $N_p-1$. There is one
coefficient in~(\ref{s3eq6}) that is not needed due to the charge
neutrality. In other words, the last coefficient can be derived from
the front $N_p-1$ coefficients, i.e., $t_{N_p}= -\sum_{i=1}^{N_p-1}
t_i$.

In the above, the inner product between two functions is defined as~\cite{ChewBook1995}
\begin{equation}\label{s3eq8}
\left\langle f_1, f_2 \right\rangle= \int f_1(\mathbf{r}) f_2(\mathbf{r}) d\mathbf{r}
\end{equation}
where the integral is assumed to converge. Moreover, the above
matrix system is irrelevant to the loop-space part of electric flux
$\textbf{D}_l$ because it is divergence free, i.e.,
$\nabla\cdot\mathbf{D}_l=0$.

\subsubsection{Impose Neumann boundary condition}
The above derivation only works for zero Neumann boundary condition, namely,
$$
\frac{\partial}{\partial n}\phi(\mathbf{r})=0 \mbox{ $\qquad$ for $\mathbf{r}\in\Gamma$}.
$$
When the normal components of the electric field is not trivial on the boundary, we need to handle it carefully.

Physically, Eq.~(\ref{s3eq2.1}) implies that Neumann boundary
condition corresponds to the 2D surface charge density introduced
into the solution system. Hence, we can add extra half RWG basis
functions at the boundary where non-zero Neumann boundary condition
appears. The coefficients of those half RWG functions are
proportional to normal electric field at corresponding boundary
points.

\subsubsection{Fast tree solver}
The matrix system (\ref{s3eq9}) can be solved with $\textit{O}(N_t)$
operations using the fast tree solver. Its key component relies on
back substitution along a tree structure (e.g., see
Fig.~\ref{fig:tree}). From a mathematical point of view, we can
invert the matrix $\overline{\mathbf{K}}$ of Eq.~(\ref{s3eq9}) with
linear complexity since a row permutation of the matrix is a lower
triangle matrix with very few elements per row. This can be achieved
with the help of topological information. We refer interested
readers to Chapter~5 in \cite{ChewBook2008} for details.

\subsection{Loop space projection}\label{S3s3}

If the media is homogeneous, both the electric flux and electric
field have no loop-space or divergence-free component existing,
namely, $\nabla\times\mathbf D=\nabla\times\mathbf E=0$. Hence, we
can project the obtained $\mathbf{D}_t$ onto the loop space or
divergence-free (solenoidal) space in order to remove the divergence
free part to retrieve the desired field. This is because the desired
electric field is orthogonal to the loop (divergence-free) space.
The procedure is called divergence-free field removal as
in~\cite{mySISC}.

Unfortunately, it is not the case in inhomogeneous media where the
electric flux has both divergence-free and curl-free parts. It is
not orthogonal to the loop space any more. However, the electric
field $\mathbf E$ is curl free and orthogonal to the divergence-free
space. Hence, we project the electric field $\mathbf{E}$ onto the
divergence-free space, in order to remove its divergence-free
component.

By solving the Eq.~(\ref{s3eq9}), coefficients $\{t_i\}_{i=1}^{N_t}$
have been obtained and known while $\{l_i\}_{i=1}^{N_l}$ are the
unknowns. We can transform Eq.~(\ref{s3eq3}) to electric field
\begin{equation}\label{s3eq14}
\mathbf{E}(\mathbf{r})=\sum_{i=1}^{N_l}
l_i\frac{\mathbf{L}_i(\mathbf{r})}{\epsilon(\mathbf{r})}+\sum_{i=1}^{N_t}
t_i\frac{\mathbf{T}_i(\mathbf{r})}{\epsilon(\mathbf{r})}.
\end{equation}
Since the electric field lives in the curl free (irrotational)
space, which is orthogonal to the loop space, we have
\begin{equation}\label{s3eq15}
\left\langle \mathbf{L}_i, \mathbf{E} \right\rangle=0.
\end{equation}
Testing Eq.~(\ref{s3eq14}) by loop basis functions and substituting the above into it, we obtain a matrix system
\begin{equation}\label{s3eq16}
\overline{\mathbf{G}}_l \cdot \mathbf{I}_l= \mathbf{V}_t
\end{equation}
where
\begin{equation}
\left[\overline{\mathbf{G}}_l \right]_{ij}=\left\langle \mathbf{L}_i(\mathbf{r}), \frac{\mathbf{L}_j(\mathbf{r})}{\epsilon(\mathbf{r})}\right\rangle
\end{equation}
\begin{equation}
\mathbf{V}_t = -\overline{\mathbf{R}}\cdot\mathbf{I}_t
\end{equation}
\begin{equation} \left[\overline{\mathbf{R}}
\right]_{ij}=\left\langle \mathbf{L}_i(\mathbf{r}),
\frac{\mathbf{T}_j(\mathbf{r})}{\epsilon(\mathbf{r})}\right\rangle
\end{equation}
 $\mathbf{I}_l=[l_1,\;l_2,\;l_3,\;\cdots,\;l_{N_l}]^T$ is the unknown, and
$\mathbf{I}_t=[l_1,\;l_2,\;l_3,\;\cdots,\;l_{N_t}]^T$ is known.

Many iterative solvers could be chosen to solve Eq.~(\ref{s3eq16}),
such as CG, Bi-CGSTAB \cite{BiCGSTAB} and GMRES
\cite{SaadBook,gmres1986}. According to our experience, GMRES has
the best performance if no preconditioning technique is applied.
Once $\mathbf{I}_l$ is found, the numerically approximated curl-free
$\mathbf E$ field is known.  From it, we can derive the potential
$\phi$.

\subsection{Solution of potential }\label{S3s4}

According to the electrostatic theory, the potential satisfies
\begin{equation}\label{s3eq19}
-\nabla\phi(\mathbf{r})= \mathbf{E(r)}.
\end{equation}
We can further expand it by a set of pulse functions
\begin{equation}\label{s3eq19.1}
\phi(\mathbf{r})=\sum_{i=1}^{N_p}\nu_i P_i(\mathbf{r}).
\end{equation}

By substituting Eqs.~(\ref{s3eq14}) and  (\ref{s3eq19.1}) into Eq.~(\ref{s3eq19}) and testing it with a set of pulse basis functions, it gives
\begin{equation}\label{s3eq20}
\overline{\mathbf{K}}^T \cdot \textbf{I}_\phi= \textbf{V}_\phi
\end{equation}
where
\begin{equation}
\left[\textbf{V}_\phi \right]_{i}=- \left\langle \textbf{T}_i(\mathbf{r}) \cdot \frac{\textbf{D}(\mathbf{r})}{\epsilon(\mathbf{r})} \right\rangle
\end{equation}
\begin{equation}
\left[\overline{\bf K}^T \right]_{ij}= \left\langle
P_j(\mathbf{r}), \nabla \cdot
\textbf{T}_i(\mathbf{r})\right\rangle=\left[\overline{\bf K}
\right]_{ji}.
\end{equation}
and $\textbf{I}_\phi=[\nu_1\;\nu_2\;\nu_3\;\cdots\;\nu_{N_p-1}]^T$.
In the same manner as Section~\ref{S3s2}, there is one coefficient
in~(\ref{s3eq19.1}) that is dependent on the others. Such
coefficient has to be specified as a particular value before
Eq.~(\ref{s3eq20}) is solved. This is because the solution of the
Neumann problem is found up to an arbitrary constant, which
corresponds to a reference potential physically. This reference
potential does not affect the solution of the pertinent problem.
This is also reflective of the fact that the inverse of the gradient
operator is not unique.

It is interesting that the resulting matrix is just a transpose of
the matrix $\overline{\mathbf{K}}$ from Section~\ref{S3s2}. Hence,
we can solve (\ref{s3eq20}) with the fast tree solver in
$\textit{O}(N_t)$ operations similar to Section~\ref{S3s2}.

\section{Special treatment for Dirichlet boundary}\label{Sec:S4}
Handling Dirichlet boundary conditions is of importance since
Poisson problems with Dirichlet boundary conditions commonly occur
in practical applications. As discussed in above section, our
proposed Poisson solver works for Neumann problems. When it comes to
the Dirichlet problems or mixed boundary problems, however, we need
to transform the original problem to the Neumann problem to
approximate it numerically.

If we consider the Poisson problem as shown in Fig.~\ref{fig:PE}, then the governing equation is
\begin{equation}\label{s4eq0}
\renewcommand{\arraystretch}{1.5}
\begin{array}{ll}
\nabla\cdot\epsilon_r(\mathbf{r})\nabla\phi(\mathbf{r})=-\rho(\mathbf{r})/\epsilon_0 &\mbox{ for $\mathbf{r}\in\Omega$}   \\
\frac{\partial}{\partial n}\phi(\mathbf{r})=g(\mathbf{r}) &\mbox{ for $\mathbf{r}\in\Gamma_N$} \\
\phi(\mathbf{r})=V_l &\mbox{ for $\mathbf{r}\in\Gamma_{Dl}$} \\
\phi(\mathbf{r})=V_r &\mbox{ for $\mathbf{r}\in\Gamma_{Dr}$}
\end{array}
\end{equation}
where $V_l$ and $V_r$ are potential values imposed on left and right part of Dirichlet boundary, $\Gamma_{Dl}$ and $\Gamma_{Dr}$, respectively.

At the first place, we transform the above to be the superposition
of two Poisson problems with appropriate boundary conditions. The
first problem (denoted by P1) is
\begin{equation}\label{s4eq1}
\renewcommand{\arraystretch}{1.5}
\begin{array}{ll}
\nabla\cdot\epsilon_r(\mathbf{r})\nabla\phi_1(\mathbf{r})=-\rho(\mathbf{r})/\epsilon_0 &\mbox{ for $\mathbf{r}\in\Omega$} \\
\frac{\partial}{\partial n}\phi_1(\mathbf{r})=g(\mathbf{r}) &\mbox{ for $\mathbf{r}\in\Gamma_N$} \\
\phi_1(\mathbf{r})=V_l &\mbox{  for $\mathbf{r}\in\Gamma_{Dl}$} \\
\phi_1(\mathbf{r})=V'_r &\mbox{  for $\mathbf{r}\in\Gamma_{Dr}$} \\
\end{array}
\end{equation}
where $V'_r$ is a potential value arisen when we approximate this set equations.

The second one (denoted by P2) is as follows
\begin{equation}\label{s4eq6}
\renewcommand{\arraystretch}{1.5}
\begin{array}{ll}
\nabla\cdot\epsilon_r(\mathbf{r})\nabla\phi_2(\mathbf{r})=0 &\mbox{ for $\mathbf{r}\in\Omega$} \\
\frac{\partial}{\partial n}\phi_2(\mathbf{r})=0 &\mbox{  for $\mathbf{r}\in\Gamma_N$}   \\
\phi_2(\mathbf{r})=0 &\mbox{  for $\mathbf{r}\in\Gamma_{Dl}$} \\
\phi_2(\mathbf{r})=V_r-V'_r &\mbox{ for $\mathbf{r}\in\Gamma_{Dr}$}.
\end{array}
\end{equation}
Obviously, the solution of original problem is just
\begin{equation}\label{s4eq7.3}
\phi(\mathbf{r})=\phi_1(\mathbf{r})+\phi_2(\mathbf{r}).
\end{equation}
Although these two sets of equations satisfy the Dirichlet boundary
condition, we can resolve them by finding the solutions of two
Neumann problems, as described in the following.

\subsection{Solution of P1}\label{S4s1}
\begin{figure}[h]
\centerline{\includegraphics[width=0.4\columnwidth,draft=false]{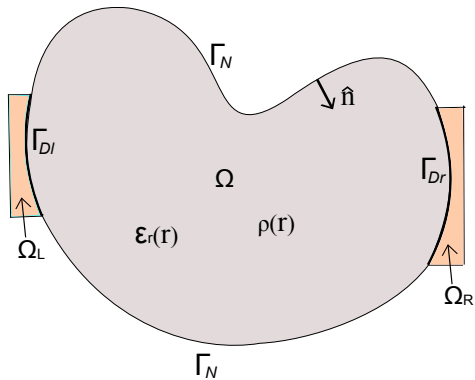}}
\caption{A schema for treatment of Dirichlet boundary conditions:
Two extended regions, $\Omega_L$ and $\Omega_R$, are introduced and
assumed filled with extremely high permittivity material.}
\label{fig:DBC1}
\end{figure}
From electromagnetic theory, the surface of a perfect electric
conductor (PEC) or perfect magnetic conductor (PMC) serves as
equipotential surface under static or quasistatic conditions. Hence,
a Dirichlet boundary condition occurs when the equipotential value
is known. In addition, it is known that a PEC may be regarded as a
dielectric material with infinite permittivity.


Based on the above theory, a technique has been developed to
surmount the difficulties arising from Dirichlet boundary
conditions. Its basic scheme is illustrated in Fig.~\ref{fig:DBC1}:
Two small regions, namely, the left extended region $\Omega_L$ and
the right extended one $\Omega_R$, are introduced into the solution
system; both regions are filled with enormously high permittivity
dielectric materials that mimic PECs to achieve equipotential
surfaces. In the context of numerical implementation, we normally
endow these materials with an enormously large relative
permittivity, e.g., $10^6$.

Since the electric field inside high permittivity material becomes
weak, and the electric field is related to the gradient of potential
($\mathbf{E}=-\nabla\phi$), we can reasonably assume that the
boundary of extended regions, $\Gamma'_D$, has homogeneous Neumann
boundary condition, where
$$\Gamma'_D=(\partial\Omega_L-\Gamma_{Dl})+(\partial\Omega_R-\Gamma_{Dr})$$
as illustrated in Fig. \ref{fig:DBC1}.

As a result, we obtain the following equation subject only to the
Neumann boundary condition to resolve problem P1:
\begin{equation}\label{s4eq8}
\renewcommand{\arraystretch}{1.5}
\begin{array}{ll}
\nabla\cdot\epsilon_r(\mathbf{r})\nabla\phi_1(\mathbf{r})=-\rho_1(\mathbf{r})/\epsilon_0 &\mbox{  for $\mathbf{r}\in\Omega$}\\
\frac{\partial}{\partial n}\phi_1(\mathbf{r})=0 &\mbox{ for $\mathbf{r}\in\Gamma'_D$} \\
\frac{\partial}{\partial n}\phi_1(\mathbf{r})=g(\mathbf{r}) &\mbox{ for $\mathbf{r}\in\Gamma_N$}.
\end{array}
\end{equation}
It should be mentioned that, the charge distribution
$\rho_1(\mathbf{r})$ has identical distribution as the original
$\rho(\mathbf{r})$ in $\Omega$, whereas the extended regions, i.e.,
$\Omega_L$ and $\Omega_R$,  need to be treated carefully as below.

At first, in accordance to (\ref{s3eq2.1}),  $\rho_1(\mathbf{r})$ must satisfy
\begin{equation}\label{s4eq11}
\int_{\Omega_L+\Omega_R} \rho_1(\mathbf{r}) d\mathbf{r}=
\int_{\Gamma_N} \epsilon(\mathbf{r}) g(\mathbf{r}) d\textit{l}-
\int_{\Omega} \rho(\mathbf{r}) d\mathbf{r}=Q',
\end{equation}
where $Q'$ is just a constant for a particular problem. It implies
the quantity of charge for which the solution should compensate to
guarantee charge neutrality. To this end, a simple but effective way
is to allocate a line charge with the quantity of $Q'$ inside either
$\Omega_L$ or $\Omega_R$.

Next, with appropriate setting of $\rho_1(\mathbf{r})$ in extended
regions, we can solve Eq.~(\ref{s4eq8}) by our new method in
Section~\ref{Sec:S3}. It should be noted that, when solving the
above equations, we must adjust the reference potential so that
$\phi_1(\mathbf{r})$ has the value of $V_l$ on the left extended
region $\Omega_L$. This can be done because the Neumann problem
needs a specified a reference potential point, as discussed in
Section~\ref{S3s4}, and high permittivity material yields an
equipotential on the surface. Hence, we can adjust that reference
potential for convenience. In the right extended region $\Omega_R$,
by contrast, a floating potential $V'_r$, that is not necessarily
equal to $V_r$, will appear (we call it floating potential because
this equipotential value is unknown before the solution is
obtained). The arising of this floating potential does not matter
for our method since we can compensate it in the following
procedure.

\subsection{Solution of P2}
\begin{figure}[h]
\centerline{\includegraphics[width=0.4\columnwidth,draft=false]{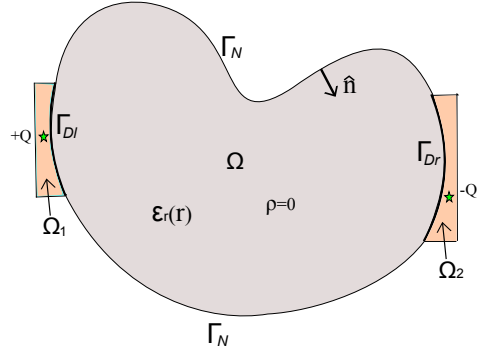}}
\caption{A schema for solving P2: Two extended regions, $\Omega_L$
and $\Omega_R$, are introduced and assumed filled with extremely
high permittivity material; two line sources that have identical
quantity but opposite signs are allocated in $\Omega_L$ and
$\Omega_R$ respectively.} \label{fig:DBC2}
\end{figure}

We have shown in the above section 
that the Dirichlet problem P1 can be solved by a Neumann problem
with extended high permittivity regions. This technique can be
applied to solving the second problem P2 as well. The solution of P2
is obtained by solving
\begin{equation}\label{s4eq12}
\renewcommand{\arraystretch}{1.5}
\begin{array}{ll}
\nabla\cdot\epsilon_r(\mathbf{r})\nabla\phi_2(\mathbf{r})=Q[\delta(\mathbf{r}-\mathbf{r}_1)-\delta(\mathbf{r}-\mathbf{r}_2)]
&\mbox{  $\mathbf{r}\in(\Omega+\Omega_L+\Omega_R)$} \\
\frac{\partial}{\partial n}\phi_2(\mathbf{r})=0 &\mbox{  for
$\mathbf{r}\in\Gamma'_D$ or $\Gamma_N$}
\end{array}
\end{equation}
where $\delta(\mathbf{r})$ refers to a 2-dimensional Dirac delta
function that corresponds to a line charge and $Q$ is a value
relative to the potential difference $V_r-V_r'$ between $\Omega_L$
and $\Omega_R$. Moreover, $\mathbf{r}_1$ is one point in $\Omega_L$
while $\mathbf{r}_2$ is another point of $\Omega_R$. This scheme is
illustrated in Fig.~\ref{fig:DBC2}.

A numerical difficulty, however, arises in the above equations
because the value of $Q$ is not determined at this point.
Fortunately, this value of $Q$ is unimportant and can be ignored in
the solution process. To this end, instead of solving
Eq.~(\ref{s4eq12}) directly, we first find the solution of another
Poisson problem with unit line charge:
\begin{equation}\label{s4eq14}
\renewcommand{\arraystretch}{1.5}
\begin{array}{ll}
\nabla\cdot\epsilon_r(\mathbf{r})\nabla\tilde{\phi}_2(\mathbf{r})=\delta(\mathbf{r}-\mathbf{r}_1)-\delta(\mathbf{r}-\mathbf{r}_2),
&\mbox{  $\mathbf{r}\in(\Omega+\Omega_L+\Omega_R)$} \\
\frac{\partial}{\partial n}\tilde{\phi}_2(\mathbf{r})=0, &\mbox{ for
$\mathbf{r}\in\Gamma'_D$ or $\Gamma_N$}.
\end{array}
\end{equation}

Next, having obtained the solution $\tilde{\phi}_2$, we can examine
the potential difference between two extended regions, $\Omega_L$
and $\Omega_R$. More specifically, suppose this potential difference
is $\tilde{V}_2$, we have
\begin{equation}
\phi_2(\mathbf{r})=\frac{V_r-V_r'}{\tilde{V}_2}\tilde{\phi}_2(\mathbf{r}).
\end{equation}
This is because for Eqs.~(\ref{s4eq12}) and (\ref{s4eq14}) the
potential difference between two ends are linearly proportional to
the quantity of charge in question.

Clearly, the solution process of Eq.~(\ref{s4eq14}) has no
difference from that of P1. In addition, since we need to solve this
equation only once for a particular geometry, the computational load
of this procedure is small in many applications, such as the
electron transport problem in which Poisson equations are solved
repeatedly.

\section{Numerical results}\label{Sec:S5}
In this section, several examples of 2-D Poisson problems will be
solved to demonstrate the effectiveness of the proposed method.
First, two typical 2-D Poisson problems will be simulated to show
the validity and efficiency. Then, this solver will be applied to a
Poisson problem that arises from a practical double-gate MOSFET
electron transport simulation.

This new algorithm has been implemented in C++ and compiled with an
Intel compiler. Moreover, all simulations listed below are performed
on an ordinary PC with $2.66$ GHz CPU, $4$ GB memory and Windows
operating system. Furthermore, in order to give a fair comparison,
the same sparse matrix structure is used for both the proposed
Poisson solver (denote by PPS below) and the finite element method
(FEM).

\subsection{Simple heterogeneous Poisson problem}

First, we consider a 2-D Poisson equation in heterogeneous media as a reference problem:
\begin{equation}
\begin{array}{cl}
\nabla\cdot\epsilon_r(x,y)\nabla\phi(x,y)=-\pi \cos(\pi x)-\pi \cos(\pi y) & \mbox{ $\qquad$ $(x,y)\in\Omega$}
\end{array}
\end{equation}
where $\Omega = \left[ 0,1 \right] \times \left[ 0,1 \right]$, and the relative permittivity
$$
\epsilon_r=\left\{
\begin{array}{cl}
1, & x<0.5 \\
2, & x\geq0.5.
\end{array}\right.
$$

For convenience, we assume that this problem has zero Neumann
boundary condition and a reference potential $2/\pi$ is imposed at
the origin.

\begin{figure}[h]
  \centering
  \subfigure{
    \includegraphics[width=0.4\columnwidth]{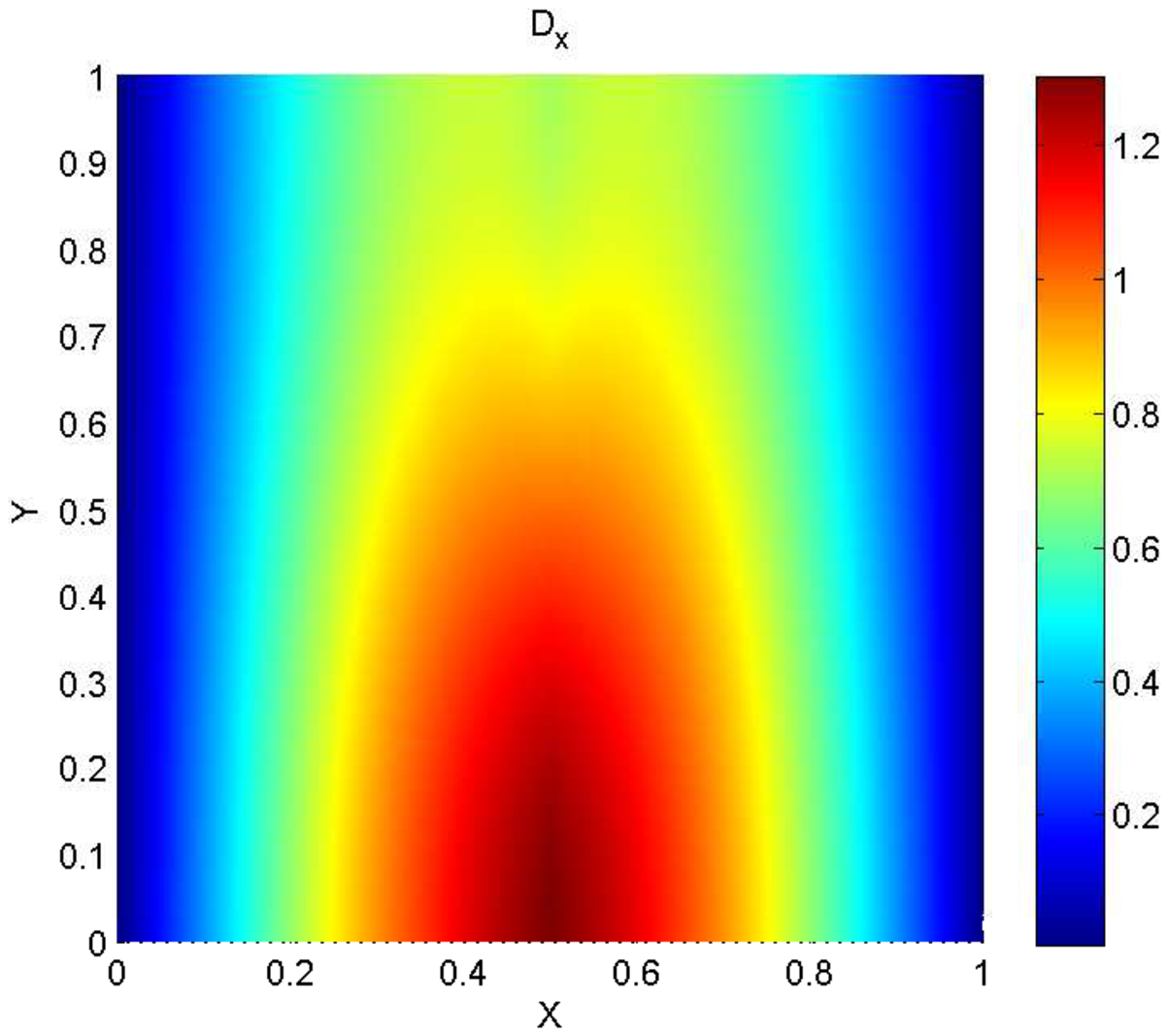}}
  \hspace{0.4in}
  \subfigure{
    \includegraphics[width=0.4\columnwidth]{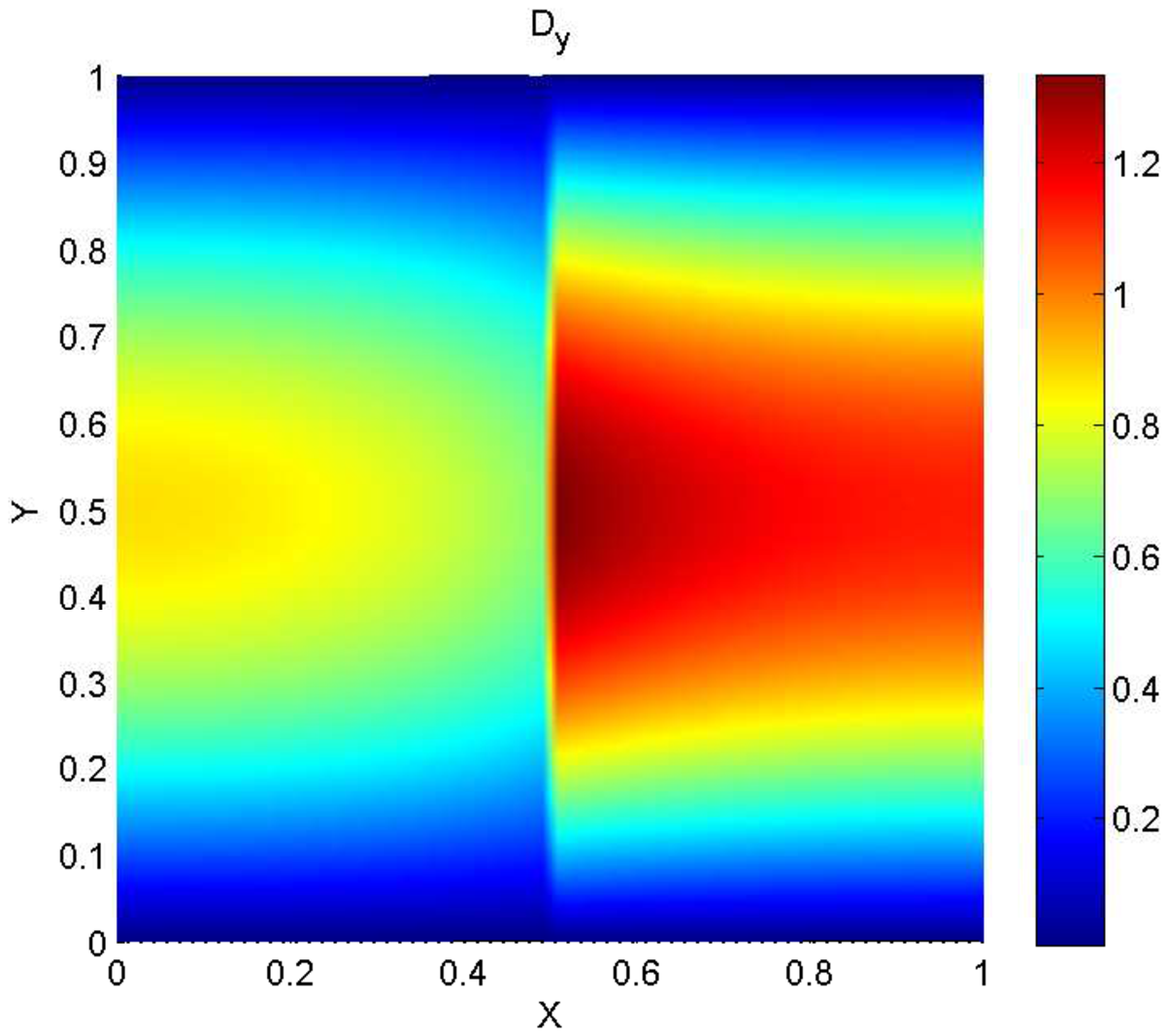}}
  \caption{The electric electric flux density calculated by the proposed method.
  Left: $x$ component of the electric flux density $\mathbf{D}$. Right: $y$ component of
  the electric
  flux density
  $\mathbf{D}$.}
  \label{fig:ex1fld} 
\end{figure}

\begin{figure}[h]
  \centering
  \subfigure{
    \includegraphics[width=0.4\columnwidth]{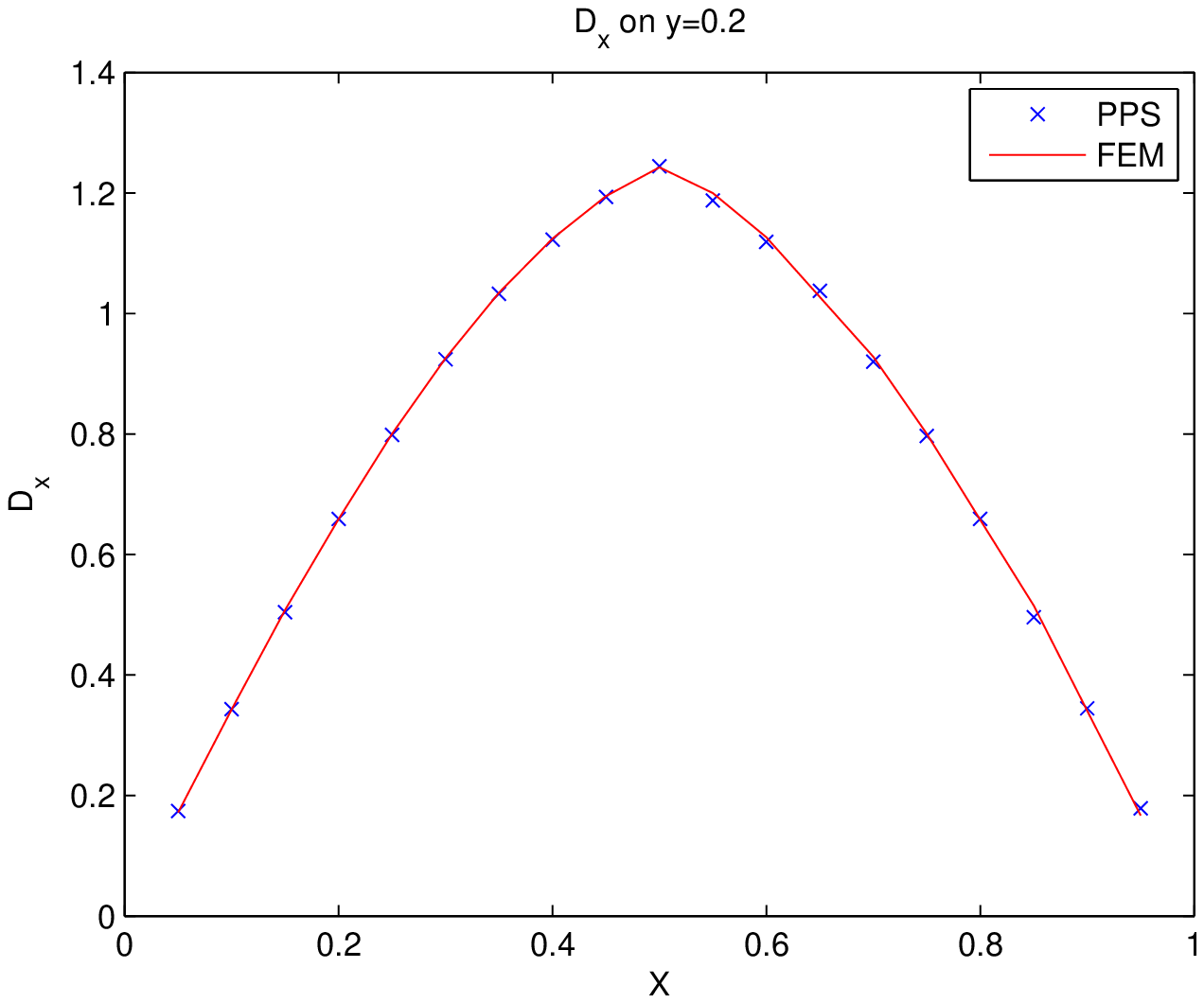}}
  \hspace{0.4in}
  \subfigure{
    \includegraphics[width=0.4\columnwidth]{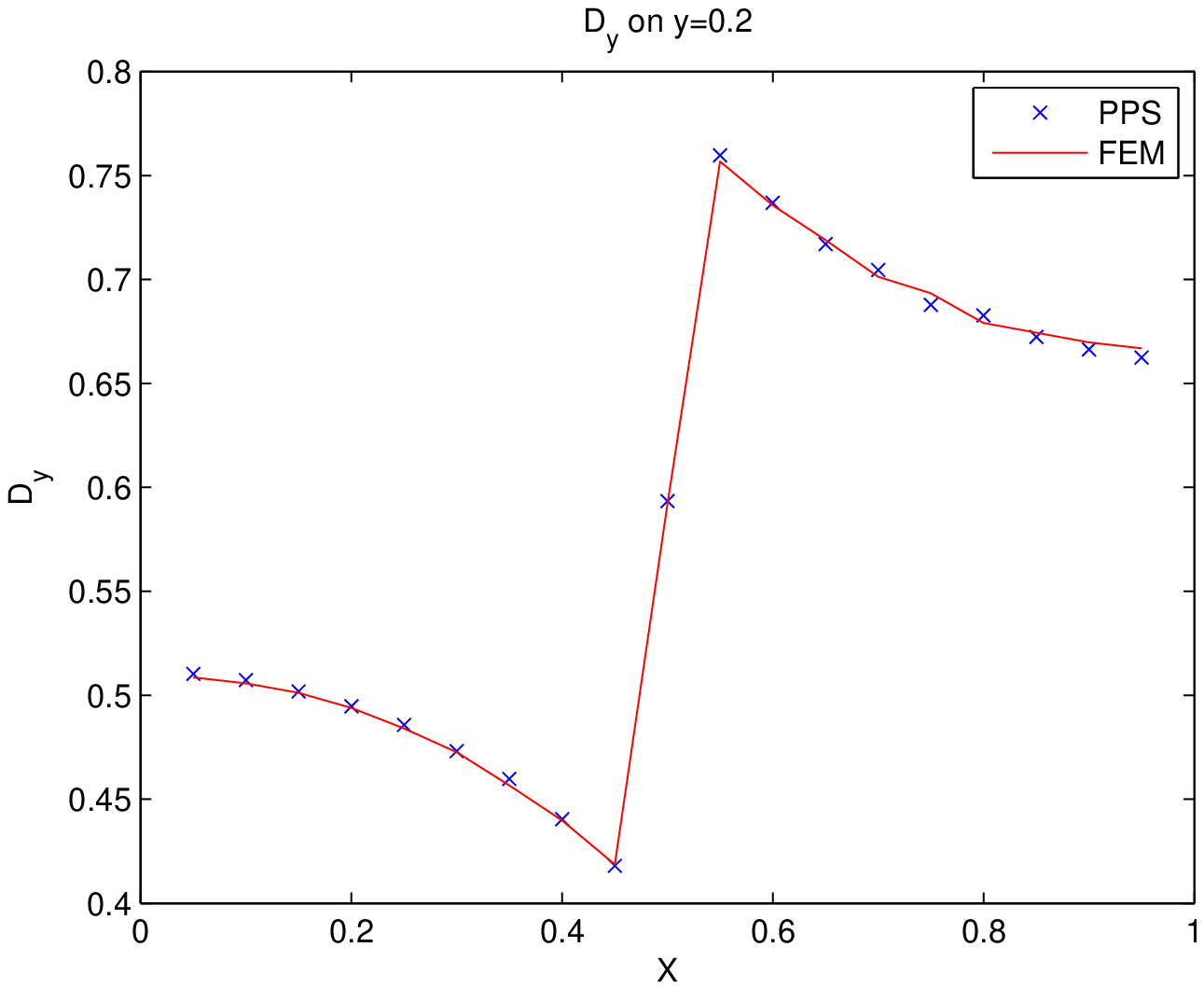}}
  \caption{Comparison of the electric flux, $\mathbf{D}$, along one line between result of proposed Poisson solver (PPS)
  and that of FEM. Left: $x$ component of the electric flux. Right: $y$ component
  of the
  electric
  flux.}
  \label{fig:ex1cprd} 
\end{figure}

\begin{figure}[h]
  \centering
  \subfigure{
    \includegraphics[width=0.4\columnwidth]{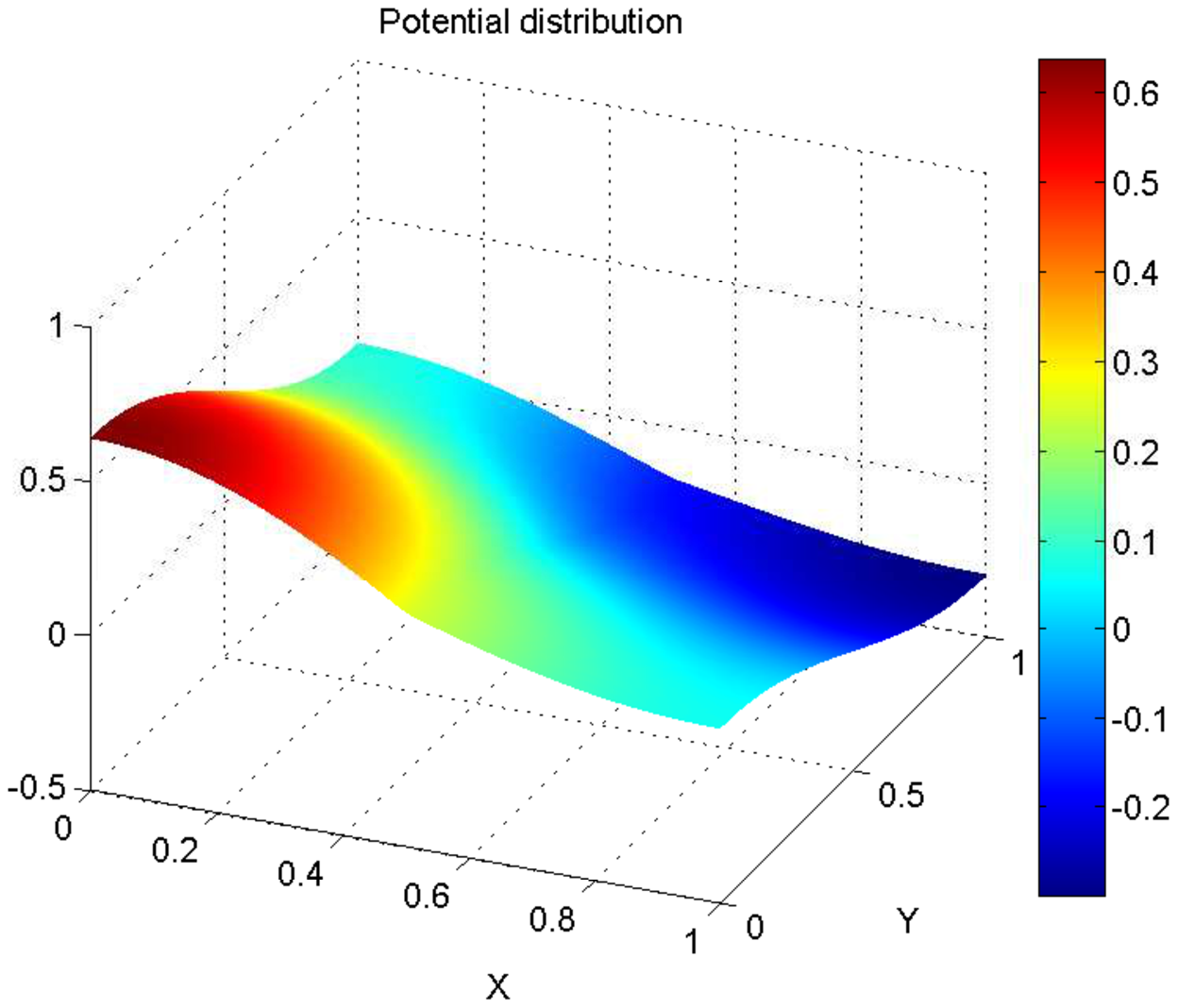}}
  \hspace{0.4in}
  \subfigure{
    \includegraphics[width=0.4\columnwidth]{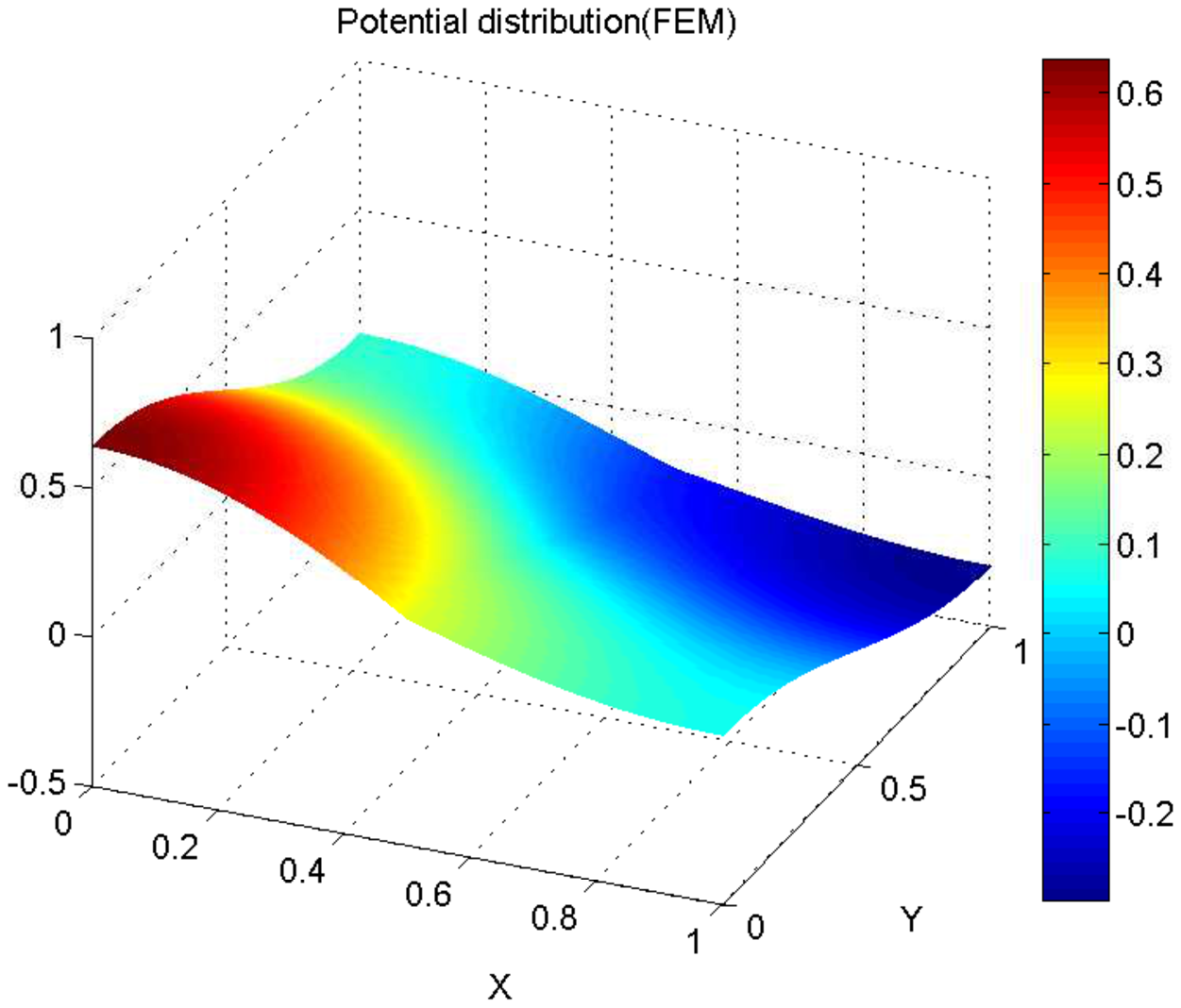}}
  \caption{Calculated potential distribution. Left: the result of the proposed method. Right: the result of FEM.}
  \label{fig:ex1potl} 
\end{figure}

\begin{figure}[h]
  \centering
  \subfigure{
    \includegraphics[width=0.4\columnwidth]{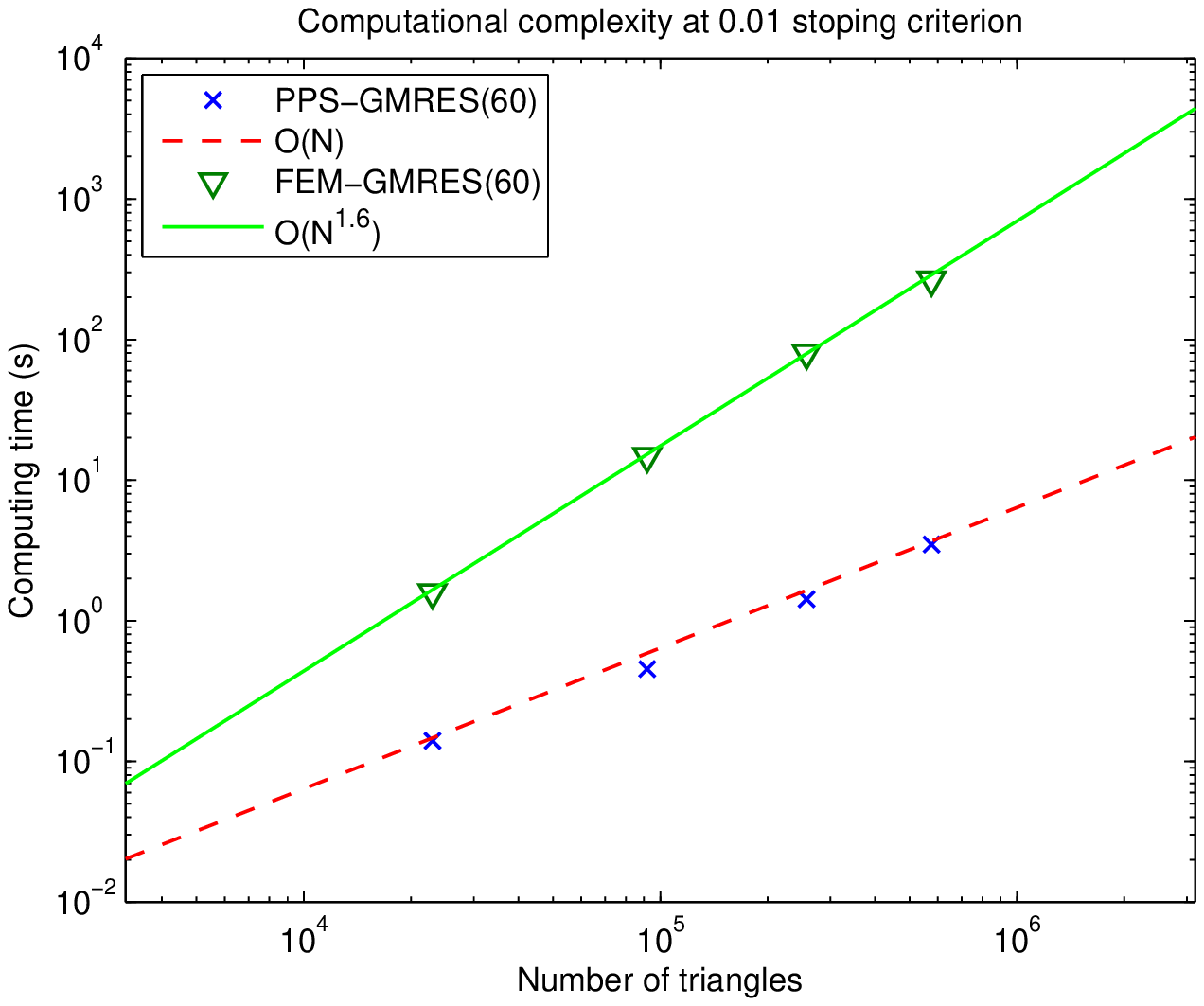}}
  \hspace{0.4in}
  \subfigure{
    \includegraphics[width=0.4\columnwidth]{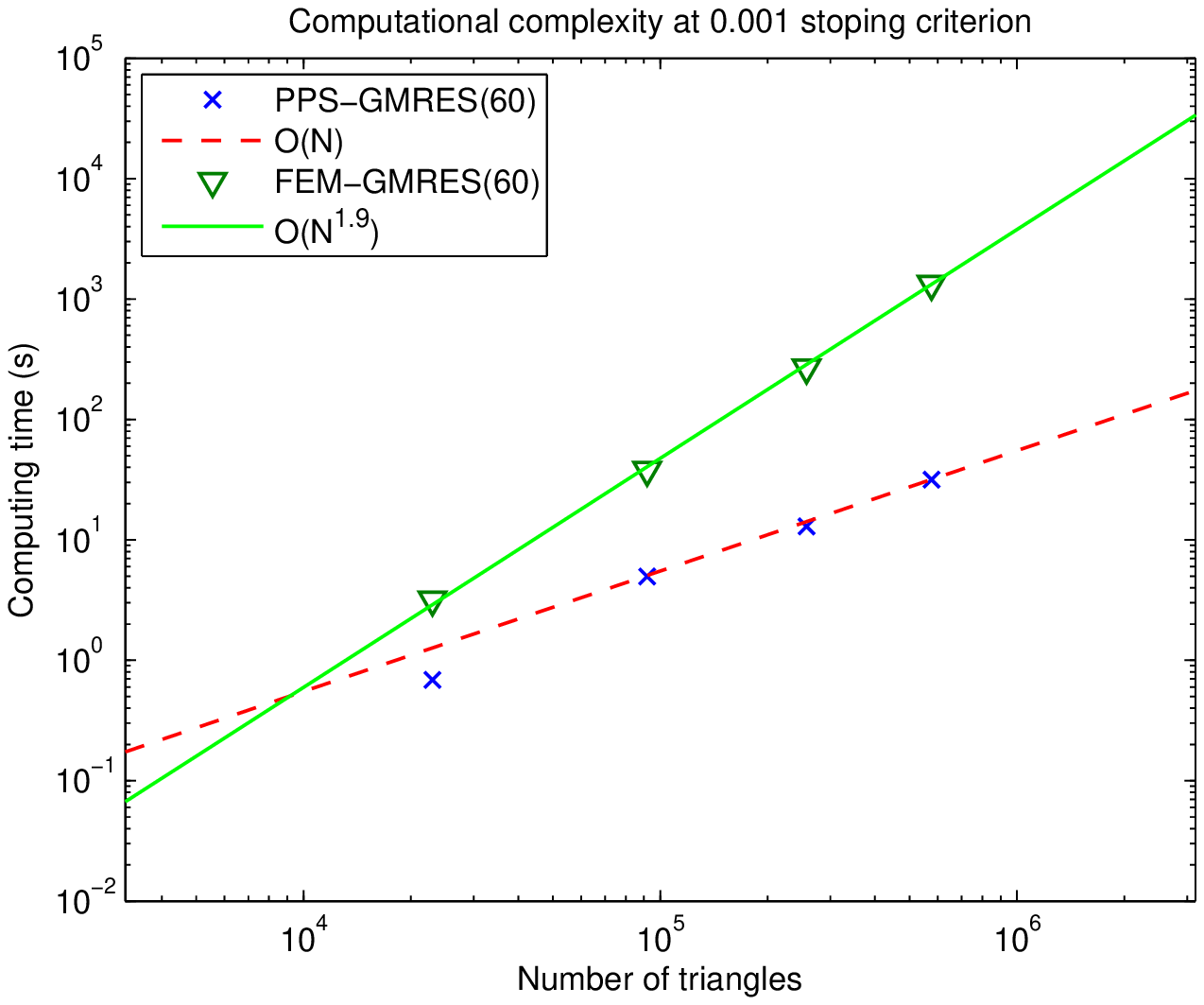}}
  \caption{Computational complexity comparison between the proposed Poisson solver (PPS) and FEM method. Left: 0.01 stopping criterion. Right: 0.001 stopping criterion.}
  \label{fig:ex1cpx} 
\end{figure}

Fig.~\ref{fig:ex1fld} shows calculated $x$- and $y$-components of
electric flux density $\mathbf{D}$. Because there is discontinuity
for $\epsilon_r$ at $x=0.5$, $y$-components, $D_y$, appears as an
abrupt change in the middle correspondingly, which is in complete
agreement with the fundamental theory of electromagnetism.
Therefore, our proposed method has the capability to handle
discontinuous media. In addition, further comparisons between
electric fluxes obtained from proposed method and traditional finite
element method are given in Fig.~\ref{fig:ex1cprd}, from which we
can see a good agreement at every sample points. Moreover,
Fig.~\ref{fig:ex1potl} shows calculated potential distributions, in
which the result of proposed method is given in the left figure
while the FEM one is given in the right figure as a reference.

To examine the computational complexity, we plot the total computing
time as the mesh density increases (see Fig.~\ref{fig:ex1cpx}). This
computing time includes three parts: (a) the first fast-tree
solution time corresponding to Section~\ref{S3s2}, (b) loop-space
projection time corresponding to Section~\ref{S3s3} and (c) the
second fast tree solution time corresponding to Section~\ref{S3s4}.
Among them, (a) and (c) can be computed expeditiously with veracious
$\textit{O}(N)$ complexity which has been proved
in~\cite{JSZhao2000}, whereas (b) includes an iterative procedure
that dominates the total solution time. Apparently, the iteration
solution time depends on the iterative solver type and the required
accuracy level. In our numerical experiment, we observe that the
total solution time of our proposed method is close to
$\textit{O}(N)$ complexity when GMRES solvers, without any
preconditioning techniques, are applied in this problem with two
different stopping criterion: $1\times10^{-2}$ and $1\times10^{-3}$.
By contrast, as shown in plots of Fig.~\ref{fig:ex1cpx}, the
computational complexities of traditional FEM method are
approximately of  $\textit{O}(N^{1.5})$ and $\textit{O}(N^{1.9})$ at
the given stopping criterion, respectively. Hence, this new proposed
method demonstrates impressive efficiency compared with the FEM
method.

\subsection{Complex Poisson problem with Dirichlet boundary condition}
\begin{figure}[h]
\centerline{\includegraphics[width=0.45\columnwidth,draft=false]{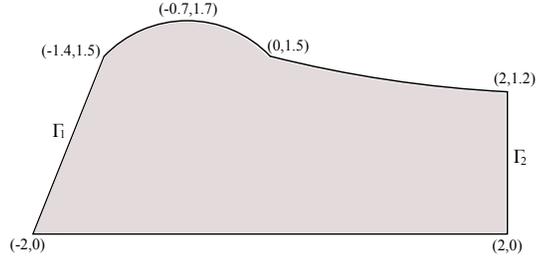}}
\caption{A two dimensional region where the Poisson problem is
defined.} \label{fig:flagshape}
\end{figure}

Next, to show the capability of handling Dirichlet boundary
conditions, we simulate a general two-dimensional Poisson equation,
which is
\begin{equation}\label{s5eq2}
\begin{array}{cl}
\nabla\cdot\epsilon_r(\mathbf{r})\nabla\phi(\mathbf{r})=-\delta(\mathbf{r}-\mathbf{r}') &\mbox{ $\qquad$ for $\mathbf{r}\in\Omega$}\\
\end{array}
\end{equation}
with boundary condition
$$
\renewcommand{\arraystretch}{1.5}
\begin{array}{ll}
\phi(\mathbf{r})=1.0 &\mbox{ $\mathbf{r}\in\Gamma_1$} \\
\phi(\mathbf{r})=0.8 &\mbox{ $\mathbf{r}\in\Gamma_2$} \\
\frac{\partial}{\partial n}\phi(\mathbf{r})=0 &\mbox{ $\mathbf{r}\in$ other boundaries,}
\end{array}
$$
and $\mathbf{r}'$ is the point $(-0.2,0.6)$.

Fig.~\ref{fig:flagshape} shows the specifications of solution region
$\Omega$. This problem is excited by a line source that is located
at point $\mathbf{r}'$ with unit charge and imposed Dirichlet
boundary conditions on the left and right edges.

\begin{figure}[h]
  \centering
  \subfigure{
    \includegraphics[width=0.4\columnwidth]{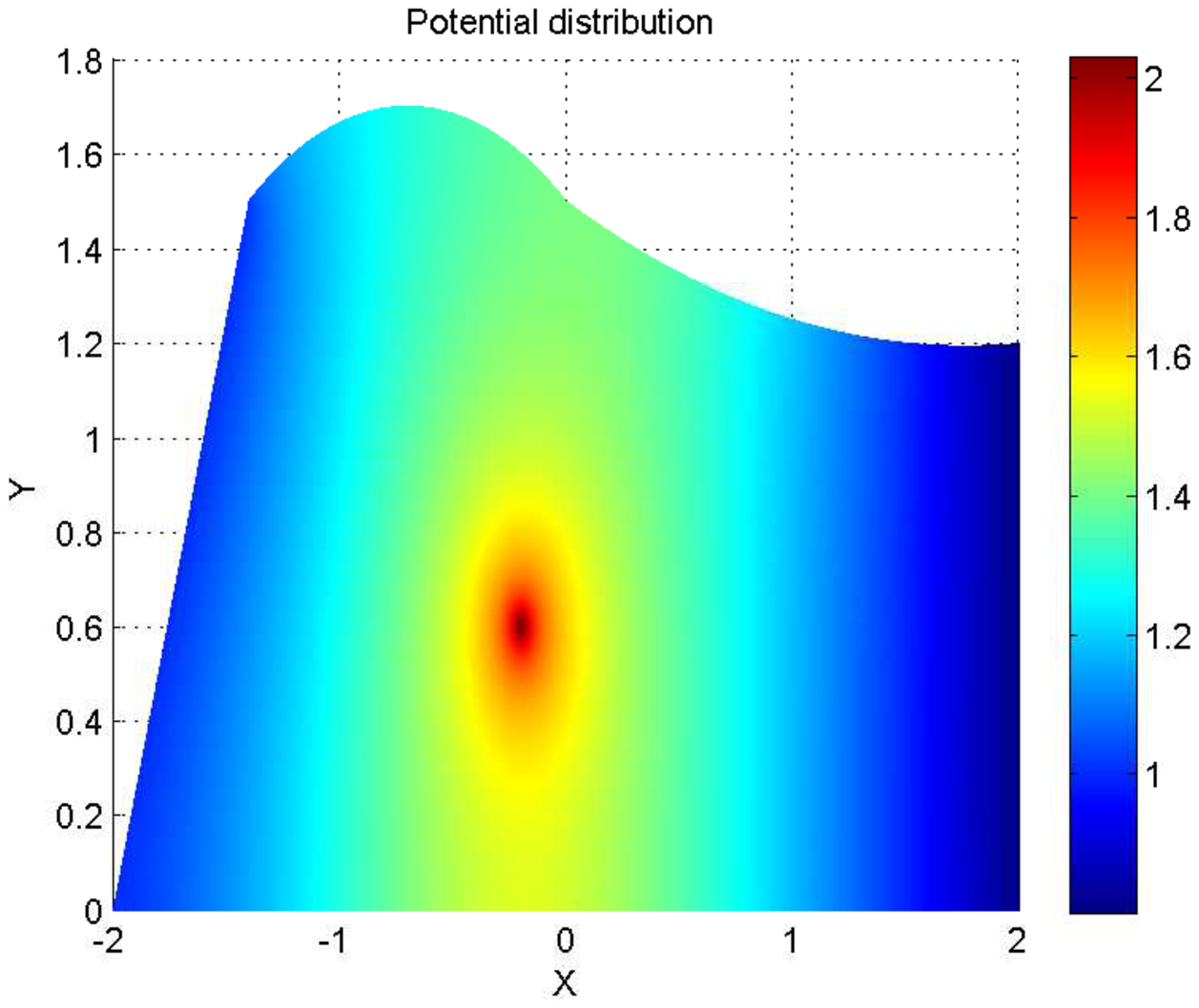}}
  \hspace{0.4in}
  \subfigure{
    \includegraphics[width=0.4\columnwidth]{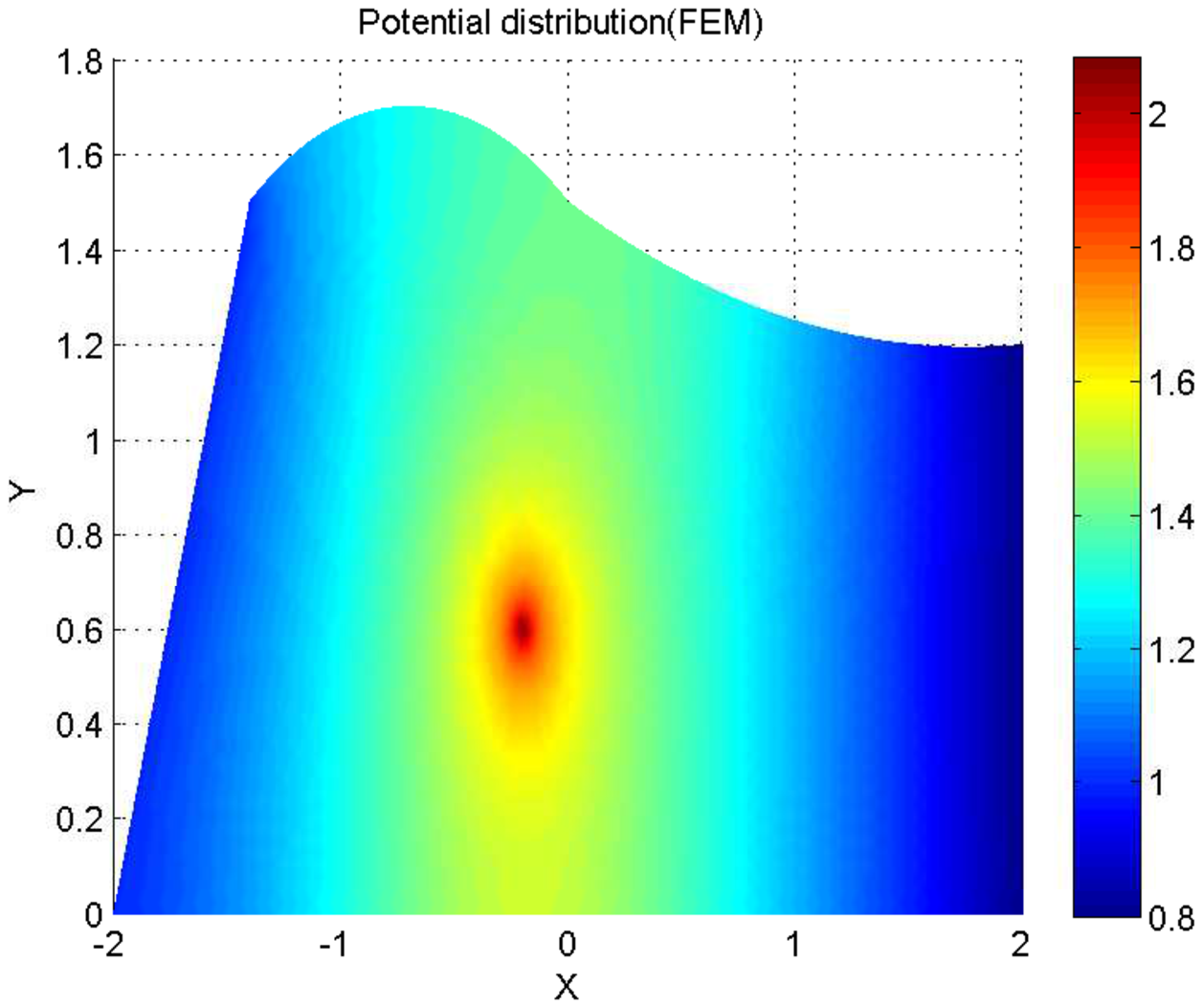}}
  \caption{Potential distribution calculated from two method. Left: the proposed method. Right: FEM method.}
  \label{fig:ex2pot} 
\end{figure}

With the technique described in Section~\ref{Sec:S4}, we can solve
this problem by the proposed method. In our numerical experiment,
the region is discretized into $133,388$ triangle patches, using the
following function to approximate the line source
$$
\delta(x,y)=\left\{
\renewcommand{\arraystretch}{1.5}
\begin{array}{cl}
625 & |x|<0.02, |y|<0.02 \\
0   & \mbox{otherwise}.
\end{array}\right.
$$
As for the iterative solver, GMRES with restart number $60$ is
employed. Finally, the calculated potential distribution is shown in
Fig.~\ref{fig:ex2pot}: The left figure shows the potential
distribution calculated by our proposed method while the right one
gives FEM results as the reference. It is shown that our proposed
method can produce the same results as the reference one. Hence, not
only Neumann boundary conditions, but also Dirichlet boundary
conditions can be handled by our proposed solver competently.

To inspect the efficiency, we observe the pertinent computing
resource of both the proposed method and the traditional FEM method.
Both methods apply GMRES iterative solver with a restart number
$60$, without any preconditioning. Table~\ref{tab1} lists the
solution time and iteration number for both methods. As seen from
this table, both iteration number and solution time of the proposed
method are considerably less than that of the traditional FEM
method.

\begin{table}\label{tab1}
\begin{center}
  \caption{A comparison of solution times between proposed method and traditional finite element method.}
  \begin{tabular}{ |c | c  c | c  c| }
  \hline
         & Proposed method & & FEM & \\ \hline
   Stopping criteria  & Iterative steps & Solution times (second)  & Iterative steps & Solution times (second)\\ \hline
   $1\times10^{-2}$   &     26          &      0.779            &    31           &    0.889         \\
   $1\times10^{-3}$   &     104         &      4.742            &    256         &     11.216        \\
   $1\times10^{-4}$   &     295         &     14.446            &    2468         &     111.976        \\
   $1\times10^{-5}$   &     593         &     28.89             &    4813        &      297.539      \\
  \hline
  \end{tabular}
\end{center}
\end{table}

\subsection{Application to double-gate MOSFET}

\begin{figure}[h]
\centerline{\includegraphics[width=0.7\columnwidth,draft=false]{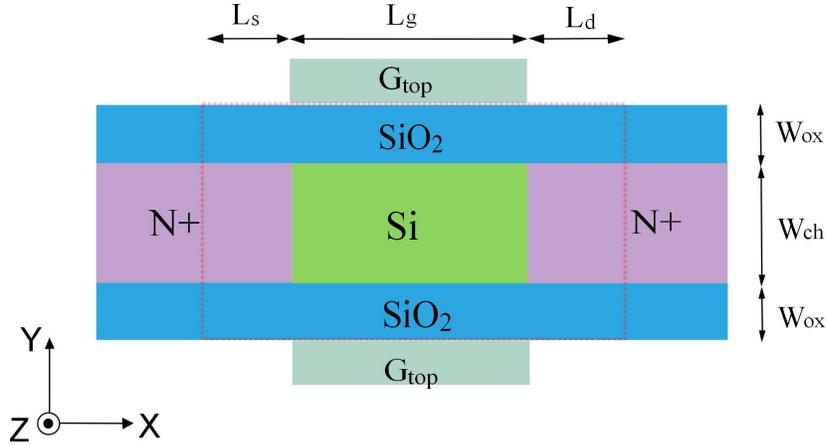}}
\caption{Two-dimensional view of the n-type double-gate silicon MOSFET.}
\label{fig:transistor}
\end{figure}

Finally, the solver is used to simulate a multigate silicon MOSFET,
as shown in Fig.~\ref{fig:transistor}, which is a promising
candidate for the next generation nanotransistor. The structure is
infinite in the $z$ direction. Gate length is denoted by $L_g$;
source and drain extension lengths are denoted by $L_s$ and $L_d$,
respectively; silicon channel thickness is $W_{ch}$; and oxide
thickness is $W_{ox}$. In our simulation, the device parameters are
$L_g=10$ nm, $L_s=L_d=4$ nm, $W_{ch}=5$ nm, and $W_{ox}=1$ nm;
doping density is $N^+=10^{26}/m^3$; the relative permittivity of
silicon is $11.9$; and relative permittivity of the silicon dioxide
is $3.8$. Other parameters can be found in~\cite{HuangChew2012}.

In this simulation, we find the solution of the following Poisson equation:
\begin{equation}\label{S5eq5}
\nabla\cdot\left[\epsilon(x,y)\nabla V_D(x,y)\right]= q[n(x,y)-N_d(x,y)]
\end{equation}
where $V_D$ is the potential to be found, $n(x,y)$ is electron
density distribution given by Schr\"{o}dinger solver. Here, $N_d$ is
the doping density and $q$ is the electron charge which is equal to
$1.62\times10^{-19}$ coulombs. The simulation domain is the region
enclosed by dashed line in Fig.~\ref{fig:transistor}. Moreover, the
Dirichlet boundary condition is enforced at the gate region, whereas
the floating boundary condition, i.e., the zero normal derivative,
is applied at the remaining boundary.

\begin{figure}[h]
\centerline{\includegraphics[width=0.7\columnwidth,draft=false]{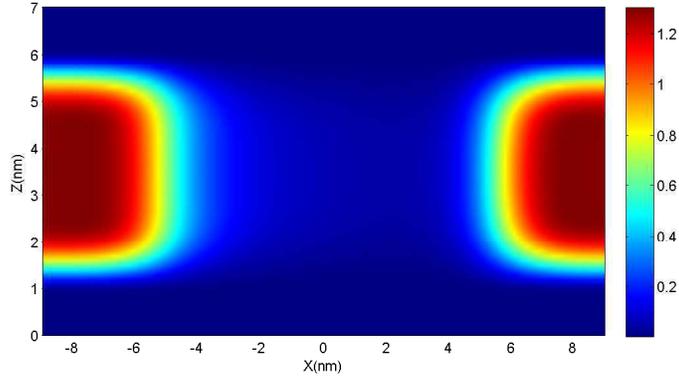}}
\caption{Two-dimensional plot of the electron density distribution.}
\label{fig:ex3c}
\end{figure}

\begin{figure}[h]
\centerline{\includegraphics[width=0.7\columnwidth,draft=false]{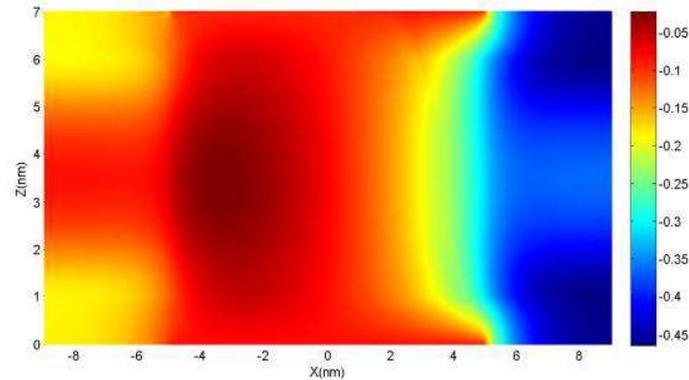}}
\caption{Two-dimensional plot of calculated potential distribution.}
\label{fig:ex3potl}
\end{figure}

Fig.~\ref{fig:ex3c} demonstrates the electron density distribution
provided from Schr\"{o}dinger solver. We then use this electron
density distribution as the source of Eq.~(\ref{S5eq5}). By our
proposed method, the potential distribution is the same as the
result in~\cite{HuangChew2012} and is shown in
Fig.~\ref{fig:ex3potl}.

\section{Conclusions}
\label{Sec:S6}

We have proposed and developed an efficient numerical method for
solving Poisson problems. It can deal with Poisson equation with
both Neumann and Dirichlet boundary conditions in a 2-D irregular
region filled with homogeneous or inhomogeneous dielectrics. In this
method, the electric flux is first found as opposed to traditional
methods. This can be done by quasi-Helmholtz decomposition with
loop-tree basis functions whose coefficients could be gained rapidly
with fast solvers. Although the Dirichlet boundary condition appears
as an numerical obstacle for the proposed new method, it can be
overcome using special treatment: small extended regions with
extremely high permittivity material are introduced to force the
potential to be constant value to imitate the pertinent Dirichlet
boundary conditions. Through numerical examples, it is shown that
this new method can solve a general Poisson equation arising from
electrostatics robustly, and has better performance than the
traditional finite element method. According to our numerical
experiments, it is observed that the computational complexity of
this new method is close to $\textit{O}(N)$.    This method is a
novel fast Poisson solver that can serves as a feasible alternative
to multigrid scheme for Poisson solutions.

\section*{Acknowledgements}
This work was supported in part by the Research Grants Council of
Hong Kong (GRF 711609, 711508, 711511 and 713011), in part by the
University Grants Council of Hong Kong (Contract No. AoE/P-04/08)
and HKU small project funding (201007176196).

The authors would like to thank Dr. Min Tang, Jun Huang and Dr. Yumao Wu for their helpful discussions and modeling data.

\end{document}